\documentclass[amsmath,amssymb,aps,prx,reprint,superscriptaddress,floatfix]{revtex4-2}
\usepackage{amsmath}
\usepackage[table]{xcolor}
\usepackage{amsfonts,amssymb}
\usepackage{graphicx}
\usepackage{hyperref}
\usepackage{academicons}
\usepackage{multirow}
\usepackage[version=4]{mhchem}
\usepackage{makecell}
\usepackage{pgf}
\usepackage{booktabs}
\usepackage[english]{babel}

\makeatletter

\adddialect\l@en\l@english
\makeatother

\graphicspath{{figure_new/main/}{figures/}}
\newcommand{\figinc}[3]{\IfFileExists{figure_new/main/#2}{\includegraphics[width=#1]{figure_new/main/#2}}{\includegraphics[width=#1]{figures/#3}}}

\extrafloats{200}
\begin{document}

\title{A practical route to amorphous solid-state electrolyte structures by hybrid reverse Monte Carlo with foundation potentials
}

\author{Xintian Wang}
\affiliation{Department of Materials Science and Engineering, National University of Singapore, Singapore}

\author{Xinyu Zhang}
\affiliation{Institute of Major Scientific Facilities for New Materials, Southern University of Science and Technology, Shenzhen 518055, China}

\author{Denys Butenko}
\affiliation{Institute of Major Scientific Facilities for New Materials, Southern University of Science and Technology, Shenzhen 518055, China}

\author{Bolong Hong}
\affiliation{Department of Physics \& State Key Laboratory of Quantum Functional Materials \& Guangdong Basic Research Center of Excellence for Quantum Science, Southern University of Science and Technology, Shenzhen 518055, China}
\affiliation{Shenzhen Key Laboratory of Solid State Batteries \& Guangdong Provincial Key Laboratory of Energy Materials for Electric Power, Southern University of Science and Technology, Shenzhen 518055, China}

\author{Jinlong Zhu}
\email{zhujl@sustech.edu.cn}
\affiliation{Department of Physics \& State Key Laboratory of Quantum Functional Materials \& Guangdong Basic Research Center of Excellence for Quantum Science, Southern University of Science and Technology, Shenzhen 518055, China}
\affiliation{Institute of Major Scientific Facilities for New Materials, Southern University of Science and Technology, Shenzhen 518055, China}
\affiliation{Shenzhen Key Laboratory of Solid State Batteries \& Guangdong Provincial Key Laboratory of Energy Materials for Electric Power, Southern University of Science and Technology, Shenzhen 518055, China}

\author{Peichen Zhong}
\email{zhongpc@nus.edu.sg}
\affiliation{Department of Materials Science and Engineering, National University of Singapore, Singapore}

\date{\today}

\begin{abstract}
Amorphous alkali oxychlorides are fast ion conductors, but their atomic structure remains unresolved. Total scattering sums all partial pair correlations into one curve that has no unique inversion, so the structure must be modeled. Melt-quench molecular dynamics builds a model from the interatomic potential alone, and reverse Monte Carlo (RMC) builds one from the measurement alone. The two have rarely been compared with all else held fixed. We benchmark both routes and two hybrid schemes on amorphous \ce{Li_{0.9}NbO_{0.9}Cl_{4.1}} against one neutron pair distribution function with one foundation potential. The hybrid schemes put the energy either in the acceptance rule (inline) or in short relaxations between blocks of refinement (alternating). Each refinement starts from a melt-quench glass or from a random packing of isolated units. The glass has the lowest energy and places the first-shell peaks correctly, but carries almost none of the oscillation beyond 6~\AA{} (goodness of fit $R^{2} = 0.29$). Pure RMC fits the measurement to within its counting noise ($R^{2} = 0.99$), but at more than $1.3$~eV/atom above the glass, and niobium loses about half of its first-shell chlorine. Started from the glass, both hybrid schemes reach $R^{2} = 0.92$ within $0.1$~eV/atom of its energy and keep its coordination chemistry. Structures from the two inputs fit almost equally well and differ in energy by a tenth of the RMC penalty, yet they differ in whether their polyhedra are connected. The input is therefore part of the method. We recommend quenching a glass with a foundation potential and refining it against the measurement with a hybrid scheme. The resulting structure carries both the local chemistry and the medium-range order, and it is the starting point for modeling ion transport in these conductors.
\end{abstract}

\maketitle

\section{Introduction}\label{sec:intro}
Amorphous alkali oxychlorides combine fast alkali-ion transport with a formability that densifies a pellet by cold pressing~\cite{motohashiFastSodiumIonConducting2024,huangOxygenMediatedStructural2025}. The amorphous \ce{$x$Li2O-MCl_y} electrolytes (M = Ta, Hf) reach up to $6.6$~mS/cm at 25~$^\circ$C~\cite{zhangFamilyOxychlorideAmorphous2023}, and \ce{LiMOCl4} (M = Nb, Ta) exceed 10~mS/cm as cold-pressed powders~\cite{tanaka2023oxyhalide}. The transport of a mobile ion through such a material is set by the structure it moves through~\cite{binciMechanismsAlkaliIonic2026,leifeldDisorderDrivenFast2026}, in particular by the size and connectivity of the sites along its path~\cite{adams2005bvs}. For crystalline \ce{LiNbOCl4}, \emph{ab initio} dynamics attributed the conductivity to the flexibility of isolated, orientationally disordered \ce{[NbOCl4]^-} chains~\cite{singhCriticalRoleFramework2024}. About half of \ce{LiNbOCl4} is amorphous, however, a fraction that stays constant across a stoichiometry series~\cite{newnhamCorrelationCoherenceLength2025}, and no atomistic model of this amorphous phase has been reported. Atomistic configurations have been built for other amorphous lithium oxyhalides~\cite{hussainExploringSuperionicConduction2024,binciMechanismsAlkaliIonic2026}. Such a configurational model must describe the coordination of the metal and of lithium, the connectivity of the polyhedra, and the packing beyond the first shell.

An amorphous structure cannot be read directly from the measurement, because the inversion is not unique.
Many different configurations can reproduce one measurement equally well~\cite{lebail2000combining,muller2010evaluation}. 
A configuration can fit an averaged measurement such as the pair distribution function and still misrepresent the coordination or the connectivity.
Two routes are therefore used to build the structure. Each serves one of two objectives, agreement with the measurement or physical plausibility, at the expense of the other.
The bottom-up route, melt-quench molecular dynamics (MQ below), melts a cell and cools it, and the resulting glass depends on the interatomic potential~\cite{liCoolingRateEffects2017,matsutani2024geo2}. Foundation potentials are machine learning interatomic potentials (MLIPs) pre-trained across most of the periodic table~\cite{kaplanMatPES2025,barroso-luqueOpenMaterials20242024}. They have made this route available without system-specific fitting~\cite{batatiaCrossLearningElectronic2025}, but the measurement plays no part in it. 
The MQ glass can differ from the measured sample for three reasons: (1) an MLIP trained largely on small crystalline cells may not transfer to a glass~\cite{bertaniAtomicStructureNa42025}, (2) the quench rates a simulation can afford exceed laboratory rates by orders of magnitude and, in silicate glasses, act chiefly on the medium-range structure~\cite{liCoolingRateEffects2017}, and (3) the sample studied here is ball-milled rather than melt-quenched~\cite{butenko2026lnoc}.

The top-down route, reverse Monte Carlo (RMC), moves atoms until the calculated pair distribution function matches the measured one~\cite{mcgreevy1988rmc}, and with well-chosen constraints it reveals the local arrangement of atoms~\cite{keen2020totalscattering}. 
Without such constraints, it returns the most disordered configuration consistent with the data~\cite{tucker2007rmcprofile} and inherits from its input whatever the data cannot fix~\cite{muller2010evaluation,lebail2000combining}. Constraints strong enough to prevent this impose correlations of their own~\cite{fischer2006diffraction}.
That agreement with the data and physical plausibility are distinct objectives was recognized for metallic glasses in the 1990s~\cite{mendelev1998partial}, yet the two routes have rarely been compared with everything but the route held fixed. For a material built from polyhedral units the comparison matters, because a close fit to $G(r)$ does not by itself fix how the units connect.

Hybrid reverse Monte Carlo (HRMC) addresses both objectives at once by adding the energy from an interatomic potential to the refinement. It exists in two schemes: inline and alternating.
In the inline scheme, the energy enters the acceptance rule, so that every move is scored on the data and the energy together~\cite{opletal2002hybrid,opletal2008hrmc,habchi2012licl}. 
In the alternating scheme, blocks of refinement alternate with separate relaxations, short and step-capped in force-enhanced atomic refinement~\cite{pandey2015fear,pandey2016realistic} or carried to a minimum~\cite{biswas2005ecmr}. 
Either way, the energy term needs a potential for the system at hand, which has meant an empirical potential~\cite{opletal2008hrmc}, a density-functional theory (DFT) calculation affordable only for small cells~\cite{pandey2016realistic}, or an MLIP trained for the system~\cite{cuillier2024rmcprofile,ehrhardt2026pyhrmc, nicholasStructureTopologyAmorphous2025}. 
A foundation potential removes this requirement. It requires no system-specific training and can serve as the energy reference for every route, so bottom-up, top-down, and hybrid structures can be compared on a single energy scale.
In addition, generative models conditioned on scattering or spectroscopic data have recently been proposed as a third option that replaces both the sampler and the potential with a learned structural prior~\cite {kwonSpectroscopyGuidedDiscoveryThreeDimensional2023,guoGenerativeInversionSpectroscopic2026}. 

\begin{figure*}[tbp]
\centering
\includegraphics[width=\linewidth]{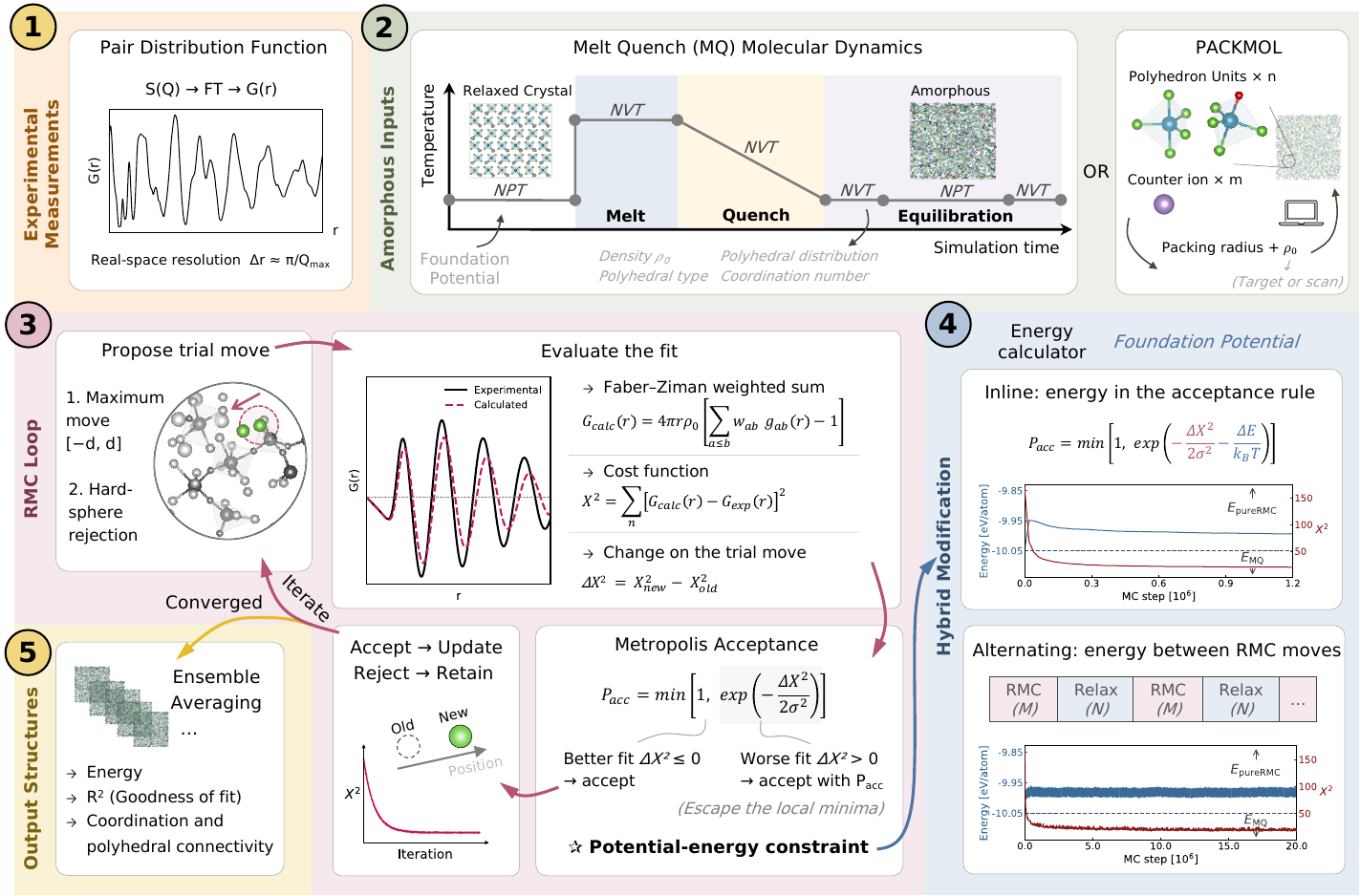}
\caption{Benchmark design. Four routes to an amorphous structure share one cell and one interatomic potential. Panel 1 shows the measurement used for the refinement tasks, a neutron total-scattering pattern transformed into $G(r)$. Panel 2 shows the two amorphous inputs, a melt-quench glass and a \textsc{packmol} packing of rigid units. Panel 3 shows the RMC loop, in which a single-atom displacement is proposed, filtered on hard-sphere cutoffs, and accepted or rejected on the change in fit. Panel 4 shows how the two hybrid schemes, inline and alternating, bring the foundation potential into the loop. Panel 5 shows the output structures, which are compared on fit, energy relative to the glass, and local chemistry. Settings are given in Methods.}
\label{fig:workflow}
\end{figure*}

Here we ask what each modeling choice, the route and the input, does to the structure it produces.
We run MQ dynamics, RMC without an energy term (pure RMC below), and both hybrid schemes (inline and alternating) on amorphous \ce{Li_{0.9}NbO_{0.9}Cl_{4.1}}~\cite{butenko2026lnoc}. 
Every route uses one neutron total-scattering curve, one cell, and one foundation potential as the energy reference (Fig.~\ref{fig:workflow}). 
We evaluate each structure on three quantities together: the fit to $G(r)$, the energy under the potential, and the niobium polyhedral chemistry. 
We report the last as the polyhedron type, the fraction of isolated Nb, and the corner-, edge-, and face-sharing fractions used for this material class~\cite{zhang2024oxyhalide,butenko2026lizrncl}.
We also start every refinement with two inputs, the melt-quench glass and a random packing of isolated units, to test whether the objectives or the inputs determine the network. 
The results single out one route: hybrid refinement of a melt-quench glass, whose structures carry both local chemistry and medium-range order.

\section{Methods}\label{sec:methods}

\subsection{Experimental reference data}\label{sec:methods:data}

The reference data are a neutron total-scattering measurement of \ce{Li_{0.9}NbO_{0.9}Cl_{4.1}}, 
a mechanochemically synthesized sample~\cite{butenko2026lnoc} in which about $55$\% of the material is amorphous (see Discussion).
Neutrons are used because they give the light elements, \ce{Li} and \ce{O}, higher scattering contrast than X-rays~\cite{keen2020totalscattering,doveReviewPairDistribution2022}. The data were collected on the Multi-Physics Instrument (BL16), a time-of-flight total-scattering diffractometer at the China Spallation Neutron Source~\cite{xuMultiPhysicsInstrument2021}. The raw time-of-flight data were corrected for detector efficiency, background, and container scattering and normalized against a vanadium standard~\cite{doveReviewPairDistribution2022}. The corrected banks were rebinned onto a common $Q$ grid, normalized, and merged to obtain the coherent total-scattering structure factor $S(Q)$. The reduced function $Q[S(Q)-1]$ was Fourier transformed over $0.5 \le Q \le 31.5$~\AA$^{-1}$ to give the real-space correlation function $G(r)$, in the neutron total-scattering formalism of Dove and Li~\cite{doveReviewPairDistribution2022}. The resulting $G(r)$ is tabulated on a $0.01$~\AA{} grid with an uncertainty of about $0.038$~\AA$^{-2}$ at each point (Fig.~\ref{fig:workflow}, panel 1).

\subsection{ML interatomic potential}\label{sec:methods:mlp}

We computed every energy in this work with one MLIP (MACE-MH-1), a multi-head foundation potential built on the message-passing atomic cluster expansion architecture (MACE)~\cite{batatiaMACEHigherOrder2023,batatiaCrossLearningElectronic2025}.
We used its MatPES r$^{2}$SCAN readout head~\cite{kaplanMatPES2025} because r$^{2}$SCAN describes formation energies, volumes, and coordination numbers of solids better than the generalized-gradient PBE level of theory~\cite{yang_rationalizing_2019,kothakondaTestingR2SCAN2023}. 
No training data specific to this composition exist, so we applied the MLIP zero-shot. 
Against plane-wave r$^{2}$SCAN calculations, the MLIP reproduces the energies with a mean absolute error of $0.66$~meV/atom for the perturbed crystal and $3.1$~meV/atom for a set of disordered $55$-atom configurations. The test set extends to $560$~meV/atom above the crystal, and energies beyond it are extrapolations (Supplementary Sec.~S1).

\subsection{Melt-quench molecular dynamics}\label{sec:methods:mqmd}

We built amorphous atomic configurations as melt-quench glasses by melting a crystalline supercell and cooling it under the MLIP. The density, coordination numbers, and polyhedron distribution are set by the MLIP rather than assumed (Fig.~\ref{fig:workflow}, panel 2). 
The starting cell is a $9\times3\times3$ supercell of the $Cc$ structure of \ce{LiNbOCl4} refined against diffraction data~\cite{singhCriticalRoleFramework2024}. 
We adjusted this supercell to the measured composition \ce{Li_{0.9}NbO_{0.9}Cl_{4.1}} by pairing each lithium vacancy with one chlorine substitution for oxygen, keeping the cell charge-balanced. This substitution yields \ce{Li292Nb324O292Cl1328} with $2236$ atoms.

\begin{table}[htbp]
\caption{Melt-quench protocol. Each phase runs in the ensemble shown at the temperature $T$ for the time $t$, and the quench in phase 3 is ramped at $5$~K/ps.}
\label{tab:mq}
\begin{ruledtabular}
\begin{tabular}{lllr}
\textrm{Phase} & \textrm{Ensemble} & \textrm{$T$ (K)} & \textrm{$t$ (ps)} \\
\colrule
1 & $NPT$ & 300 & 25 \\
2 & $NVT$ & 1800 & 200 \\
3 & $NVT$ & $1800 \rightarrow 300$ & 300 \\
4 & $NVT$ & 300 & 50 \\
5 & $NPT$ & 300 & 200 \\
6 & $NVT$ & 300 & 100 \\
\end{tabular}
\end{ruledtabular}
\end{table}

We ran the dynamics in LAMMPS~\cite{thompsonLAMMPSFlexibleSimulation2022a} using the \texttt{mliap} pair style and the protocol in Table~\ref{tab:mq}, with a timestep of $2$~fs and thermostat and barostat damping times of $0.1$~ps. We held the two constant-pressure stages at $1$~bar, and we ran every stage above room temperature at constant volume because, under constant pressure, the melt expands without bound. 
Every stage used a Nos\'e--Hoover thermostat except the quench, which used a Langevin thermostat ramped at $5$~K/ps.
Because the density of the real sample is not reliably known, we did not impose it. Instead, the second $NPT$ stage (phase 5 in Table~\ref{tab:mq}) set the density at $300$~K, and production sampling continued under $NVT$ from the frame closest to that equilibrated volume. 
We confirmed melting from the mean-squared displacement of \ce{Nb} in the melt, and the \ce{Nb} displacement plateaus after the quench (Supplementary Fig.~S2a,b). We branched five glasses from the same melt at $40$~ps intervals and quenched each with an independent seed. These five glasses serve as replicate inputs for every refinement route. They measure about $39 \times 41 \times 42$~\AA{} and span $2.00$--$2.05$~g/cm$^{3}$ ($\rho_0 = 0.0325 \pm 0.0004$~\AA$^{-3}$).

\subsection{Reverse Monte Carlo}\label{sec:methods:rmc}

RMC builds a structural model by displacing atoms 
until the pair distribution function computed from the model matches the measured one, accepting or rejecting each trial move based on how much it improves the fit rather than on energy~\cite{mcgreevy1988rmc,keen2020totalscattering}. In RMC, the number of atoms, composition, and cell remain fixed, and only the atomic positions change~\cite{mcgreevy1988rmc}. 
The loop described below is shown in Fig.~\ref{fig:workflow}, panel 3.

\paragraph{Amorphous input preparation.}\label{sec:methods:mq}
We used two inputs for each refinement: the MQ glass and a random packing, which differ in what they assume.
The MQ glass requires only a crystal and a quench protocol, whereas random packing requires specifying its building blocks in advance. 
The building blocks are rigid, isolated \ce{NbOCl4} units with an axial \ce{Nb-O} bond of $2.00$~\AA{} and \ce{Nb-Cl} bonds of $2.40$~\AA{}, and \ce{NbCl5} units with all ligands at $2.40$~\AA. These bond lengths are close to the sums of Shannon ionic radii~\cite{shannon1976radii}.\label{sec:methods:random} 
These units were mixed in the ratio $292:32$, and with $292$ free \ce{Li} this mixture reproduces the composition and the charge-balanced off-stoichiometry of the glass. We generated five such packings with \textsc{packmol}~\cite{martinez2009packmol}. In each packing, we placed the units randomly in a box of the same volume as the MQ glasses, so both inputs share the same number density.
Each niobium in a packing carries a complete ligand shell and shares no ligand with another. The two inputs are drawn side by side in Fig.~\ref{fig:workflow}, panel 2.

\paragraph{The observable.}
At each step, ten partial pair histograms $h_{ab}(r_n)$ are maintained on $M$ bins of width $\Delta r$. Dividing each by the number of pairs a uniform system with the same number density $\rho_0 = N/V$ would place in that bin yields the partial pair distribution functions $g_{ab}$~\cite{keen2020totalscattering}. 
These functions are combined into the neutron-weighted total,
\begin{equation}
\begin{aligned}
 g_{ab}(r_n) &= \frac{h_{ab}(r_n)}{\rho_0\,4\pi r_n^{2}\,\Delta r\;N_{ab}/N},\\
 G(r_n) &= 4\pi r_n\rho_0\Big[\sum_{a\le b} w_{ab}\,g_{ab}(r_n)-1\Big], \\
 w_{ab} &= \frac{c_a c_b b_a b_b (2-\delta_{ab})}{\langle b\rangle^{2}},
 \label{eq:wab}
\end{aligned}
\end{equation}
where $N_{ab}=n_a(n_a-1)$ for $a=b$ and $n_an_b$ otherwise, $c_a$ is the atom fraction, and $b_a$ is the bound coherent scattering length~\cite{sears1992lengths}. 
The weight $w_{ab}$ (i.e., the neutron weight of a pair) follows from the composition and the scattering lengths alone. 
Because $b_{\mathrm{Li}} < 0$, pairs of lithium with any other element carry a negative weight, and we quote the shares $|w_{ab}|/\sum|w_{ab}|$.
A trial move updates only the moved atom's contribution to the histograms. 
We compare the calculated and measured curves over the fitting window $1.58 \le r \le 15.0$~\AA{} in $482$ bins of width $0.028$~\AA, resampling the measured curve onto the same bins. 
The lower end is the shortest contact the hard-sphere table below permits, so that no model pair falls outside the window. The upper end lies below half the shortest perpendicular cell width, about $19.6$~\AA, so no pair in the window is counted against a periodic image~\cite{mcgreevy1988rmc,keen2020totalscattering}.

\paragraph{Metropolis sampling.}
In each step, we randomly choose one atom and displace it~\cite {mcgreevy1988rmc,tian2023llzo}, drawing each Cartesian component of the displacement from $U[-d,d]$ (Fig.~\ref{fig:workflow}, panel 3). 
The trial position is first tested against a table of pair-specific minimum-approach distances and rejected outright if it violates any~\cite{mcgreevy1988rmc,keen1990glasses,karlsson1995paddle}. Each entry is $0.95$ times the shortest distance observed for that pair in the unfitted MQ glass, and the same table serves every run whatever its input or route. 
Every route, pure RMC included, therefore inherits the closest approaches of the MQ glass as a weak constraint on ten distances.
Surviving moves are scored on $X^{2}$, the sum of squared differences between the calculated and measured curves, and accepted with the Metropolis probability~\cite{opletal2002hybrid,tian2023llzo}
\begin{align}
 X^{2} &= \sum_{n=1}^{M}\big[G(r_n)-G_{\mathrm{exp}}(r_n)\big]^{2}, \label{eq:X2}\\
 P_{\mathrm{acc}} &= \min\!\left[1,\;\exp\!\left(-\frac{\Delta X^{2}}{2\sigma^{2}}\right)\right].
 \label{eq:accept}
\end{align}
Each scheme uses only single-atom displacements, and we tested and excluded species swaps (Supplementary Sec.~S3).

\paragraph{Parameters in RMC.}
Two settings govern the RMC walk. The width $\sigma$ in Eq.~\eqref{eq:accept} sets how strongly the measurement is enforced: a smaller value makes the same $\Delta X^{2}$ less likely to be accepted. The maximum displacement $d$ sets how far the structure moves in one attempt.
The two trade off: a large $d$ with a tight $\sigma$ rejects almost everything, while a small $d$ accepts readily but explores little. 
We held $\sigma$ fixed at $0.030$~\AA$^{-2}$ in the units of $G(r)$, the value that \citet{opletal2002hybrid} used for the error of $g(r)$ and $S(Q)$, and close to the point uncertainty of $0.038$~\AA$^{-2}$ in the measured $G(r)$ (Sec.~\ref{sec:methods:data}).
For pure RMC we used $d = 1.0$~\AA{} (Supplementary Sec.~S3). We used the same $\sigma$ for both hybrid schemes, so that the measurement is enforced equally along every route.

\paragraph{Coordination-number constraint.}
A coordination-number constraint is commonly added to the objective to hold the coordination of a cation that the fit alone does not fix~\cite{hirata2016sio,matsutani2024geo2,onodera2011li2sp2s5,muller2010evaluation,drabold2009topics}.
We applied one to the \ce{Nb-O} count. The target is the distribution of \ce{O} neighbors within $2.92$~\AA{} of each \ce{Nb} in the MQ glass. 
We added the squared deviation of the model's distribution from this target, in units of a $0.05$ tolerance, to $X^{2}$ with a weight of $0.9$. We chose this weight so that the constraint balances a data term of the size used here (the term is written out in Supplementary Sec.~S4). 
Such a constraint can also bias the structure toward what its author expected~\cite{ehrhardt2026pyhrmc}, so we report pure RMC with the constraint active, and the unconstrained runs serve as a control.

\paragraph{Stopping and sampling.}
A pure RMC run stops when the running mean of $X^{2}$ changes by less than $1\%$ over three consecutive windows of twenty configurations. Sampling then continues for a further $5\times10^{6}$ moves with every setting held fixed~\cite{opletal2008hrmc,farmahini2015hrmc}, and we analyzed the final configuration of each pure RMC chain.

\subsection{Hybrid reverse Monte Carlo}\label{sec:methods:hrmc}

The MLIP enters a refinement in one of two ways. 
In the \emph{inline} scheme, the energy is added to the acceptance rule~\cite{opletal2008hrmc,habchi2012licl}, so that every trial move is scored on the measurement and on the energy at once (Fig.~\ref{fig:workflow}, panel 4, upper). In the \emph{alternating} scheme, blocks of ordinary refinement alternate with short relaxations, and the acceptance rule remains unchanged (panel 4, lower). 
Both keep the observable, the trial move, and the hard-sphere filter of pure RMC.

\paragraph{Inline.}
Acceptance follows \citet{opletal2002hybrid},
\begin{equation}
 P_{\mathrm{acc}} = \min\!\left[1,\;
 \exp\!\left(-\frac{\Delta X^{2}}{2\sigma^{2}} - \frac{\Delta E}{k_{\mathrm{B}}T}\right)\right],
 \label{eq:hrmc-accept}
\end{equation}
where $E$ is the total potential energy of the cell. 
The energy is evaluated for every move that survives the hard-sphere filter, so the MLIP evaluations set the cost of the scheme. Because the moves are symmetric and $\sigma$ and $T$ are held fixed, the rule satisfies detailed balance. An inline chain therefore samples the distribution proportional to $\exp(-X^{2}/2\sigma^{2} - E/k_{\mathrm{B}}T)$ over the configurations the hard-sphere table allows. 
The displacement is $d = 0.3$~\AA, which is smaller than in pure RMC because large moves are rarely accepted once the energy enters the rule. 
The temperature $T$ sets the exchange rate between the two terms. It is a balance parameter rather than the sample temperature~\cite{farmahini2015hrmc,opletal2008hrmc}, and we set it to $175$~K (Supplementary Sec.~S5). 
Each chain runs a fixed budget of $1.2\times10^{6}$ generated moves rather than stopping under the pure RMC rule, because the energy falls slowly under the energy term (Supplementary Sec.~S5).

\paragraph{Alternating.}
The alternating scheme follows the force-enhanced atomic refinement of \citet{pandey2016realistic}. Blocks of $M$ accepted moves alternate with $N$ steps of the fast inertial relaxation engine~\cite{bitzekStructuralRelaxationMade2006}, which replaces the conjugate-gradient steps of the original method. We used the published values $(M,N) = (500,5)$~\cite{pandey2016realistic} (Supplementary Sec.~S5). The displacement is $d = 0.15$~\AA, smaller again than in the inline scheme, because the relaxation must keep pace with the disturbance the moves cause. Each chain likewise runs a fixed budget of $2\times10^{7}$ generated moves, and neither scheme applies relaxation at the end of any run.

\subsection{Evaluation}\label{sec:methods:metrics}

Structures are compared on the goodness of fit to the measurement, on the energy, and on the local chemistry around niobium. 
The goodness of fit is the coefficient of determination over the fitting window,
\begin{equation}
 R^{2} = 1 -
 \frac{\sum_{n}\big[G(r_n)-G_{\mathrm{exp}}(r_n)\big]^{2}}
 {\sum_{n}\big[G_{\mathrm{exp}}(r_n)-\overline{G_{\mathrm{exp}}}\big]^{2}}.
 \label{eq:r2}
\end{equation}
Because each histogram bin holds a finite number of pairs, a calculated $G(r)$ carries counting noise of its own. Averaging several configurations smooths this noise and raises $R^{2}$ without changing the structure, so we quote $R^{2}$ for single configurations throughout.
Poisson resampling of a fixed configuration of $2236$ atoms sets the ceiling that this noise places on $R^{2}$ at $0.972$, which we call the counting-noise ceiling. A structure above it has fitted part of its own counting noise to the data, so structures above the ceiling are treated as equally converged rather than ranked against one another.
The energy is the potential energy, in eV/atom, relative to the MQ glass, $\Delta E = E - E_{\mathrm{MQ}}$, where $E_{\mathrm{MQ}}$ is the mean over the five independent MQ glasses. 
We averaged alternating values over the last quarter of each run, because the fit and energy oscillate with the refinement--relaxation cycle.

The local chemistry around \ce{Nb} is described by two measures. 
First, the coordination environment of each \ce{Nb} is assigned by ChemEnv~\cite{waroquiers2020chemenv} in pymatgen~\cite{ongPythonMaterialsGenomics2013}, which identifies the polyhedron the site most closely resembles from its Voronoi-derived neighbor set. 
Second, the connectivity between polyhedra is obtained from the ligands they share in the same ChemEnv assignment. A link is a pair of Nb that share at least one ligand, and links are split into corner, edge, and face sharing~\cite{zhang2024oxyhalide,butenko2026lizrncl}. 
We report the number of links per \ce{Nb} and the fraction of \ce{Nb} that share no ligand, which we call isolated.
Supplementary Table~S1 lists these values beside a simpler count of shared first-shell ligands. 
Values in the text are means over the five chains of a route. Standard deviations are given in the figures and in Supplementary Table~S1.

\section{Results}\label{sec:results}

\subsection{Melt-quench dynamics and pure RMC}
\label{sec:results:mq}

We first test melt-quench dynamics and pure RMC against the measurement and the MLIP energy. 
We quenched five glasses and ran pure RMC from each glass and from each of five random packings, with and without the coordination-number constraint. 
Every variant thus has five independent RMC chains, because refinements identical in every other respect can end in different structures~\cite{funnell2013cyclohexane,muller2010evaluation,timoshenko2024exafs}.

\begin{figure}[ht]
\centering
\includegraphics[width=\linewidth]{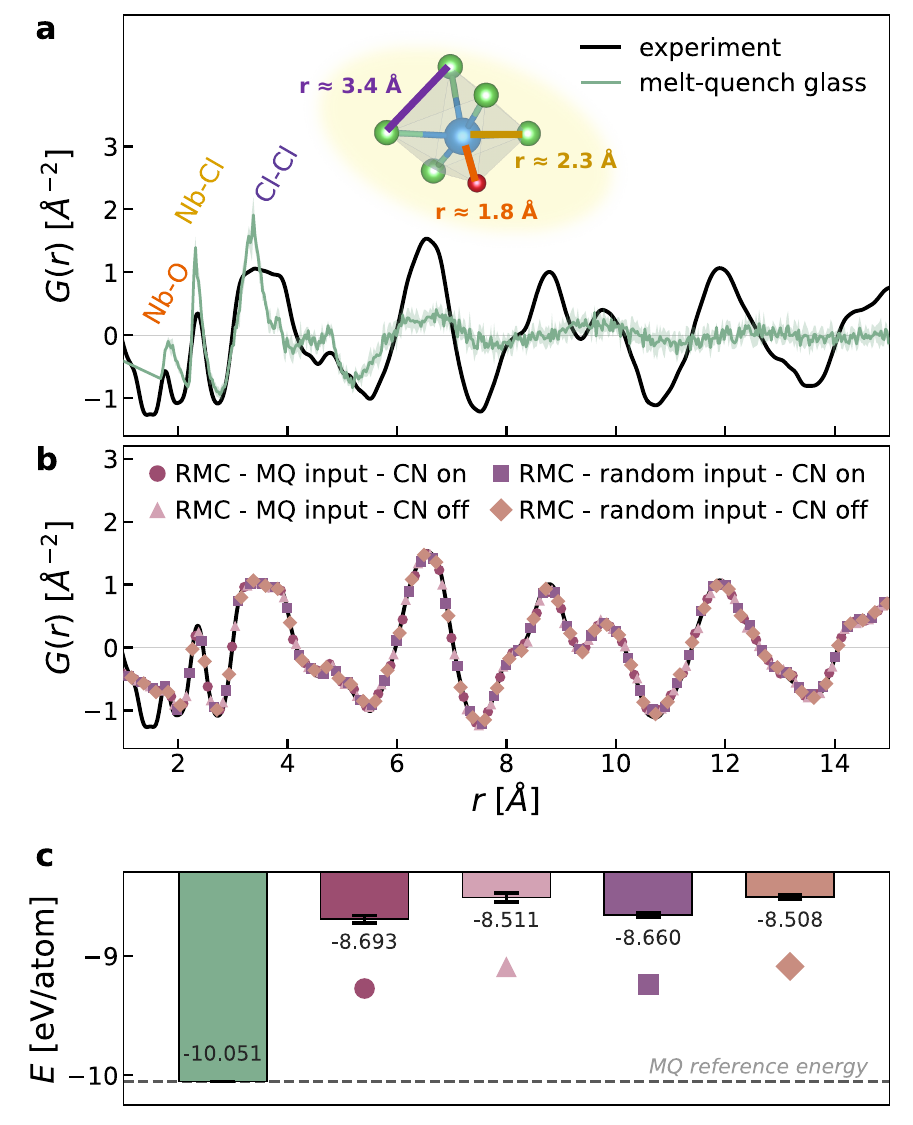}
\caption{Melt-quench glass and pure RMC against the measurement. (a) The measured $G(r)$ (black) is compared with the mean over the five melt-quench glasses (green). The green band gives one sample standard deviation. Each first-shell peak of the glass is labeled with the pair that dominates it by neutron weight, and the inset draws the three distances on one \ce{Nb} polyhedron. (b) The measurement is compared with one structure from each pure RMC variant. The legend gives the input and the coordination-number constraint (CN) setting, on or off. (c) Potential energy per atom of the glass and the same four variants, averaged over five chains with one sample standard deviation. The markers below the bars identify the variants as in (b), and the dashed line marks the energy of the melt-quench glass. In (a) and (b), curves are drawn from $1.0$~\AA{} and the fit runs from $1.58$~\AA{} (Methods).}
\label{fig:dilemma}
\end{figure}

Figure~\ref{fig:dilemma}a compares the measured $G(r)$ (black) with the mean of the five MQ glasses (green). 
A single glass configuration fits the measurement with $R^{2} = 0.29$ (Eq.~\eqref{eq:r2}).
Below about $6$~\AA{}, the MQ glass reproduces the measured peak positions to within about $0.07$~\AA. These peaks arise from distances within the \ce{Nb} polyhedron. 
The shoulder near $1.76$~\AA{} corresponds to the short \ce{Nb-O} bond, the peak near $2.36$~\AA{} to the \ce{Nb-Cl} bond, and the peak near $3.4$~\AA{} to the first \ce{Cl-Cl} shell. 
Diffraction and \emph{ab initio} molecular dynamics assign the same distances to these pairs in crystalline \ce{LiNbOCl4}~\cite{singhCriticalRoleFramework2024}.
Because the MQ glass places these distances correctly and has the lowest energy of any structure in this work (green bar in Fig.~\ref{fig:dilemma}c), we use its local chemistry as the reference for the coordination and connectivity of every other route. This reference is the MLIP prediction for this composition, not a measured structure.

The agreement is limited to peak positions below $6$~\AA. The first-shell peaks of the glass are sharper than the measured ones.
For example, the \ce{Cl-Cl} peak near $3.4$~\AA{} reaches nearly twice the measured height at a third of its width.
Beyond $6$~\AA{}, the glass carries almost none of the oscillation that the measurement holds. 
Within the quench rates we could afford, $1$--$10$~K/ps, the fit barely changes: slowing the quench tenfold lowers the glass energy by about $5$~meV/atom and raises the time-averaged $R^{2}$ of a $662$-atom glass by only $0.02$ (Supplementary Fig.~S2c).
Two other causes, which these data cannot separate, are a crystalline fraction in the ball-milled sample~\cite{butenko2026lnoc,newnhamCorrelationCoherenceLength2025} and an MLIP trained mostly on small periodic cells, which cover local bonding but not medium-range correlations (see Discussion).

\begin{figure*}[htbp]
\centering
\includegraphics[width=\linewidth]{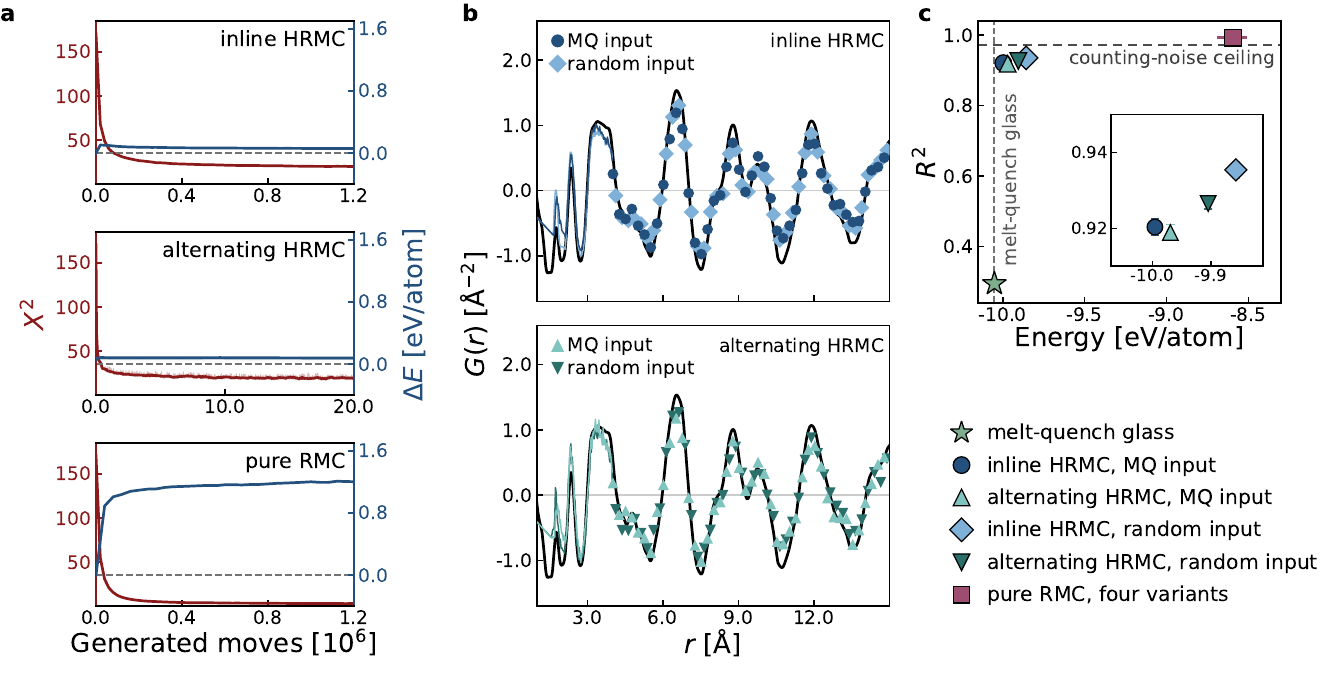}
\caption{Hybrid refinement: energy control, fit, and the trade-off between them. (a) $X^{2}$ (dark red, left axis) and the energy relative to the glass (blue, right axis) along one chain of each route from the MQ input. The alternating traces are rolling medians over the relaxation cycle, and the pure RMC energies were evaluated on saved frames after the run. (b) The measured $G(r)$ (black) is compared with one final structure from each hybrid scheme and input, drawn as a line below $4$~\AA{} and as markers above. The upper panel shows the inline scheme and the lower panel the alternating scheme. (c) Single-configuration $R^{2}$ versus potential energy per atom for each route, with the mean and one sample standard deviation over five chains. The four pure RMC variants share one marker whose bar spans their energies. The dashed lines mark the melt-quench glass energy and the counting-noise ceiling, and the inset magnifies the hybrid cluster. Alternating values are means over the last quarter of each run (Methods).}
\label{fig:pareto}
\end{figure*}

Pure RMC displaces atoms until the calculated $G(r)$ matches the measured one, in a cell whose volume the MLIP set. 
Figure~\ref{fig:dilemma}b compares the measurement with one structure from each pure RMC variant, and Fig.~\ref{fig:dilemma}c gives the energy of each variant. 
All four variants reach $R^{2} = 0.992$--$0.993$, above the counting-noise ceiling of the cell ($0.972$), but at $1.3$--$1.5$~eV/atom above the MQ glass. The constraint lowers this penalty by less than $0.2$~eV/atom (Fig.~\ref{fig:dilemma}c). Pure RMC therefore fits the measurement but yields energetically unfavorable structures under the MLIP, with or without the constraint.

\subsection{Hybrid refinement}\label{sec:results:hrmc}

To test whether the energy term lowers the energy of the refined structures without losing the fit, we applied the two hybrid schemes.
We ran both schemes (inline and alternating) on the five MQ glasses and the five random packings. 
Figure~\ref{fig:pareto}a follows $X^{2}$ (Eq.~\eqref{eq:X2}) and the energy relative to the glass along one chain of each route from the MQ input.
Under pure RMC, the energy rises by more than $1$~eV/atom within the first $2\times10^{5}$ moves and stays there, while $X^{2}$ falls below that of either hybrid scheme.
Under both hybrid schemes, $X^{2}$ falls to similar levels, and the energy stays within $0.1$~eV/atom of the glass.

The two hybrid schemes control the energy differently. 
The inline scheme evaluates the MLIP energy for every move that passes the hard-sphere filter, so it controls the energy from the first move and the energy trace is smooth.
The alternating scheme calls MLIP evaluations only during its relaxations, five steps after every five hundred accepted RMC moves. Each cycle raises the energy by about $0.025$~eV/atom, and the relaxation removes it again. 
The energy trace oscillates accordingly, and after $2\times10^{6}$ moves the chains are stationary within cycle-to-cycle scatter at $0.08$~eV/atom above the glass. 
Chains from the random input show the same two patterns (Supplementary Fig.~S4c,d).

Figure~\ref{fig:pareto}b compares the measurement with one final configuration from each hybrid scheme and input. The fit improves on the glass across the whole window.
The over-sharp first-shell peaks are brought onto the measurement, and the oscillation beyond $6$~\AA{} that the glass lacks is recovered.
Figure~\ref{fig:pareto}c plots $R^{2}$ against the energy for every route, each point the mean over five chains. The two conventional routes occupy opposite corners of the plot.
The MQ glass has the lowest energy and $R^{2} = 0.29$, whereas pure RMC reaches $R^{2} = 0.992$--$0.993$ at $1.3$--$1.5$~eV/atom above the glass. Both hybrid schemes lie in a corner neither conventional route reaches, with a close fit at low energy. From the MQ input, they reach $R^{2} = 0.92$ at $0.05$--$0.08$~eV/atom above the glass.

\begin{figure*}[htbp]
\centering
\includegraphics[width=0.9\linewidth]{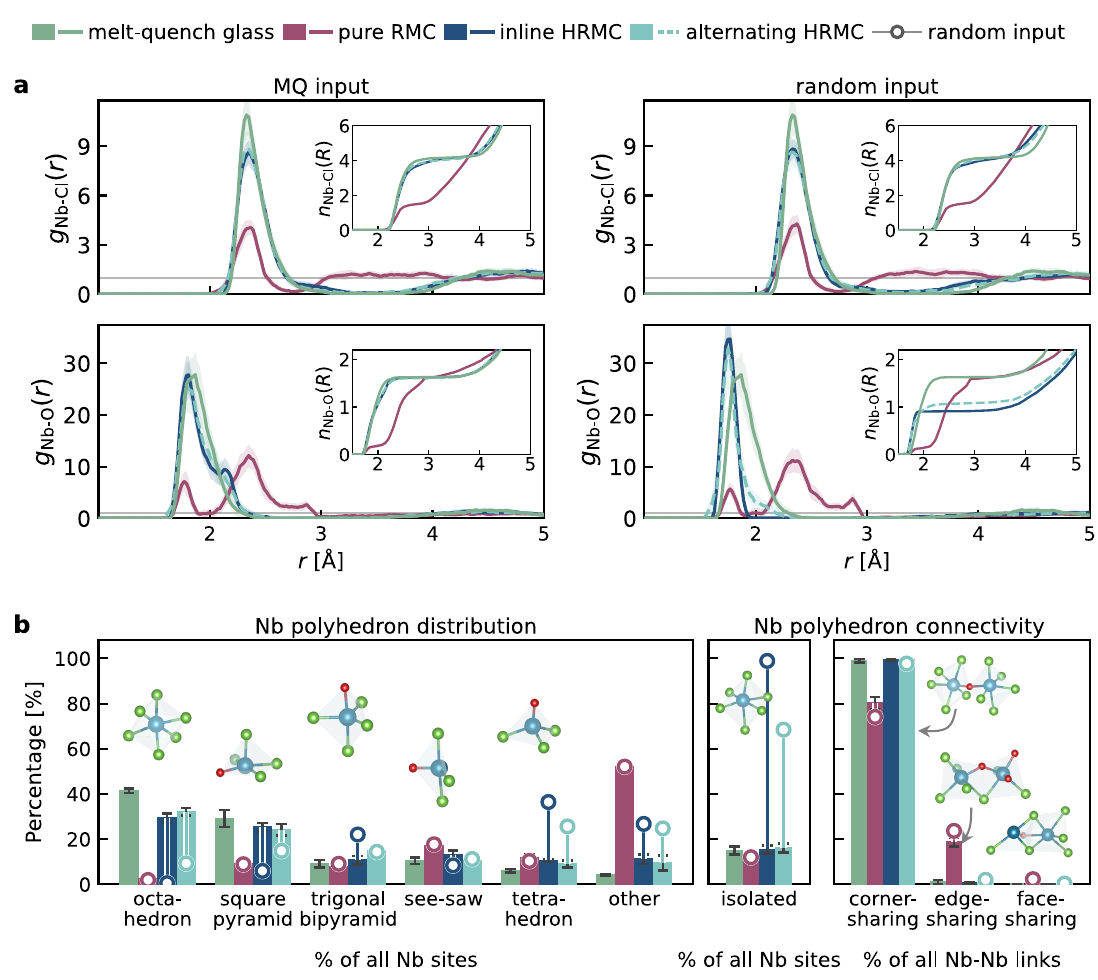}
\caption{Effect of each route on the local chemistry around \ce{Nb}. (a) \ce{Nb-Cl} (upper) and \ce{Nb-O} (lower) partial pair distribution functions from the melt-quench input (left) and the random input (right), with the melt-quench glass repeated in both columns for reference. Each curve is the mean over five chains with a band of one sample standard deviation, and pure RMC is shown with the coordination-number constraint. The insets give the running coordination number $n(R)$. Curves are drawn from $1.0$~\AA{} and the fit runs from $1.58$~\AA. (b) Distribution of \ce{Nb} coordination environments (left), fraction of isolated \ce{Nb} (center), and sharing mode of the \ce{Nb-Nb} links (right), as the mean and one sample standard deviation over five chains. Environments and the isolated fraction are percentages of \ce{Nb} sites, and the sharing modes are percentages of links. Bars give the routes from the melt-quench input, and open circles on stems give the same route from the random input. The drawings show an example of each environment and sharing mode. A circle is omitted where a route has too few links for a fraction to be defined (Supplementary Sec.~S6). The legend serves both panels.}
\label{fig:chem}
\end{figure*}

From the random input, both hybrid schemes fit slightly better than from the MQ input ($R^{2} = 0.927$--$0.935$ against $0.919$--$0.920$, each within $\pm 0.002$ between chains) but end $0.06$--$0.14$~eV/atom higher in energy. Both gaps are small besides the $1.3$~eV/atom that pure RMC adds, so the two inputs lie close together in Fig.~\ref{fig:pareto}c. Whether they also agree in structure is the question of the next section.

\subsection{Local and medium-range structure}
\label{sec:results:chemistry}\label{sec:results:medium}

We next compare the local chemistry around \ce{Nb} across the routes and the two inputs.
Figure~\ref{fig:chem}a presents the \ce{Nb-Cl} (upper) and \ce{Nb-O} (lower) partial pair distribution functions. 
From the MQ input (left), the \ce{Nb-Cl} first peak sits at the bonded distance in every route, $2.34$--$2.36$~\AA{} against $2.33$~\AA{} in the glass. 
Under pure RMC, this peak loses more than half of its height compared with the glass and returns to the uncorrelated level only beyond $3$~\AA. 

The running coordination number $n(R)$ in the insets is the number of neighbors within a distance $R$. The coordination number is its value at the first minimum of $g(r)$, where $n(R)$ plateaus. 
For the MQ glass and both hybrid schemes, the \ce{Nb-Cl} $n(R)$ plateaus near $4.1$ at the first minimum. The region between the first and second shells is empty.
For pure RMC, $n(R)$ has no plateau. It is $1.5$ at $2.8$~\AA{} and only passes $4$, the plateau value of the MQ glass, near $4$~\AA.
In the \ce{Nb-O} partial (lower), the MQ glass and both hybrid schemes share a sharp first peak at the bond length.
Under pure RMC, the coordination-number constraint holds the oxygen count, but the first \ce{Nb-O} maximum moves from $1.87$ to about $2.35$~\AA{} (Supplementary Sec.~S4). From the random input (right), the \ce{Nb-Cl} partials of both hybrid schemes match those from the MQ input. The \ce{Nb-O} running coordination, however, plateaus near $0.9$--$1.1$, close to the $0.90$ oxygen per \ce{Nb} that the packing's $292$ \ce{NbOCl4} units fix, against $1.6$ in the glass.

Figure~\ref{fig:chem}b (left) presents the distribution of Nb coordination environments. 
We show five characteristic environments with four- to six-coordinate and others. 
Under pure RMC, these five environments account for $48$\% of the sites, compared with $96$\% in the MQ glass.
Most of the remainder hold three ligands or fewer. 
Pure RMC thus strips about two \ce{Cl} from each \ce{Nb}: the running \ce{Nb-Cl} coordination within $3.20$~\AA, the first minimum of the MQ glass, falls from $4.1$ to $2.2$.
From the MQ input, both hybrid schemes cover $89$--$91$\% of their sites with the same five environments. The distribution shifts from octahedra toward the lower environments, but the running \ce{Nb-Cl} coordination stays at $4.1$, so the sites keep their ligands and are only more distorted (Supplementary Sec.~S6). From the random input (open circles), both hybrid schemes hold fewer octahedra and more tetrahedral and unmatched sites. Most unmatched sites are five-coordinate (Supplementary Sec.~S6), matching the coordination of the isolated units they started from.

Figure~\ref{fig:chem}b (center and right) gives the fraction of isolated \ce{Nb} and the sharing mode of the \ce{Nb-Nb} links. 
From the MQ input, both hybrid schemes keep the isolated fraction of the MQ glass (around $15$\% of the \ce{Nb} sites), whereas pure RMC lowers it to $11$\%. 
The random input changes this measure far more than it changes the fit or the energy (less than $0.02$ in $R^{2}$ and at most $0.14$~eV/atom, Fig.~\ref{fig:pareto}c). After inline refinement, $99$\% of the \ce{Nb} remain isolated, with $0.006$ links per \ce{Nb} against $0.72$ from the MQ input. After alternating refinement, $68$\% remain isolated, with $0.17$ links per \ce{Nb} (Supplementary Table~S1). The missing links are the most likely origin of the $0.06$--$0.14$~eV/atom by which the random-input structures lie above those from the MQ input. 
This explains the higher potential energy than initialization from the MQ glass. 

The routes also differ in how the linked polyhedra share ligands.
The glass links its polyhedra mostly through corners ($99$\% of its links), and both hybrid schemes preserve this.
Pure RMC, in contrast, has $19$\% edge-sharing links. Its links per \ce{Nb} rise from $0.74$ to $0.90$ (Supplementary Table~S1), so the edges are added to the corner-sharing network rather than converted from corners. The coordination-number constraint does not change this picture. Without it, more than half of the \ce{Nb} under pure RMC lose all their oxygen (Supplementary Sec.~S4). The constraint restores the \ce{Nb-O} count but not the bond length.
\citet{adams2005bvs} noted the same limitation for an average coordination constraint. Without the constraint, pure RMC holds $0.60$ corner-sharing and $0.10$ edge-sharing links for each \ce{Nb}, about as many links in total as the MQ glass.

\begin{figure}[htbp]
\centering
\includegraphics[width=\linewidth]{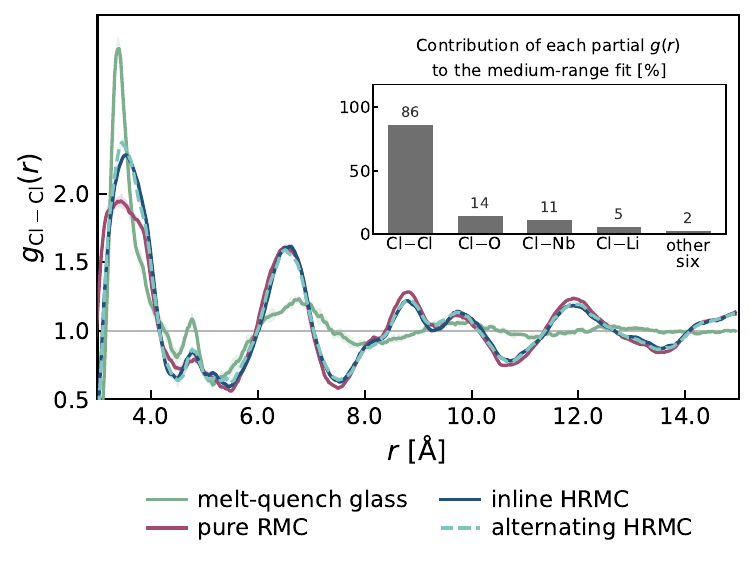}
\caption{\ce{Cl-Cl} partial pair distribution function from $3$ to $15$~\AA{} for each route from the MQ input, averaged over five chains, with pure RMC shown with the coordination-number constraint. The gray line marks $g = 1$. The inset gives, for inline refinement, the share of the improvement in the fit between $6$ and $15$~\AA{} that is recovered by replacing one partial of the glass with the refined one, for each pair in turn. The shares overlap and therefore total more than $100$\%.}
\label{fig:clcl}
\end{figure}

Figure~\ref{fig:clcl} presents the \ce{Cl-Cl} partial $g_{\text{Cl-Cl}}(r)$ from $3$ to $15$~\AA.
The inset gives the contribution of each partial to the medium-range fit, obtained by recomputing $G(r)$ of the glass with that one partial taken from the inline-refined structure. The \ce{Cl-Cl} partial alone recovers $86$\% of the improvement between $6$ and $15$~\AA, and no other pair recovers more than $14$\%.
The fit improvement over this range therefore comes almost entirely from Cl, whose pairs carry most of the neutron weight.
Beyond $6$~\AA, the \ce{Cl-Cl} partial of every refined structure oscillates with a larger amplitude and a different phase than that of the glass, and the three refined routes coincide. Each maximum marks a distance at which more \ce{Cl} pairs sit than in a uniform packing. The measurement therefore orders the chloride packing at separations that the glass leaves uncorrelated.

\section{Discussion}\label{sec:discussion}

We have benchmarked four routes to the structure of amorphous \ce{Li_{0.9}NbO_{0.9}Cl_{4.1}} against one neutron pair distribution function and one foundation potential (MACE-MH-1, r$^2$SCAN head). MQ dynamics and pure RMC each satisfy one objective, and both hybrid schemes starting from the MQ glass satisfy both, reaching $R^{2} = 0.92$ within $0.1$~eV/atom of the glass while preserving its local chemistry.

The hybrid schemes satisfy both objectives because the two terms constrain different parts of the structure.
The energy term holds the short-range geometry that $G(r)$ alone leaves underdetermined, since it sums ten partials into one curve.
Without the energy term, Nb loses about half of the Cl in its first shell under pure RMC, while the fit still exceeds the counting-noise ceiling.
In turn, the fit to $G(r)$ supplies the packing beyond $6$~\AA{} that the MQ route does not reach. In the hybrid schemes, it orders the chloride shells at these distances, while the energy term keeps the \ce{Nb-O-Nb} linkage in place.
Prior studies show that applying the two terms sequentially lets one degrade the other. Relaxation after refinement removes correlations beyond $4$~\AA{}~\cite{bousige2015mdhrmc}, and refinement after a melt-quench can raise the energy~\cite{zhao2023dataassimilation,matsutani2024geo2}, as our pure RMC runs from the MQ glass also show.
The hybrid schemes avoid this by coupling the two terms, either at every move (inline) or at short intervals (alternating).

\begin{figure*}[tbp]
\centering
\includegraphics[width=0.95\linewidth]{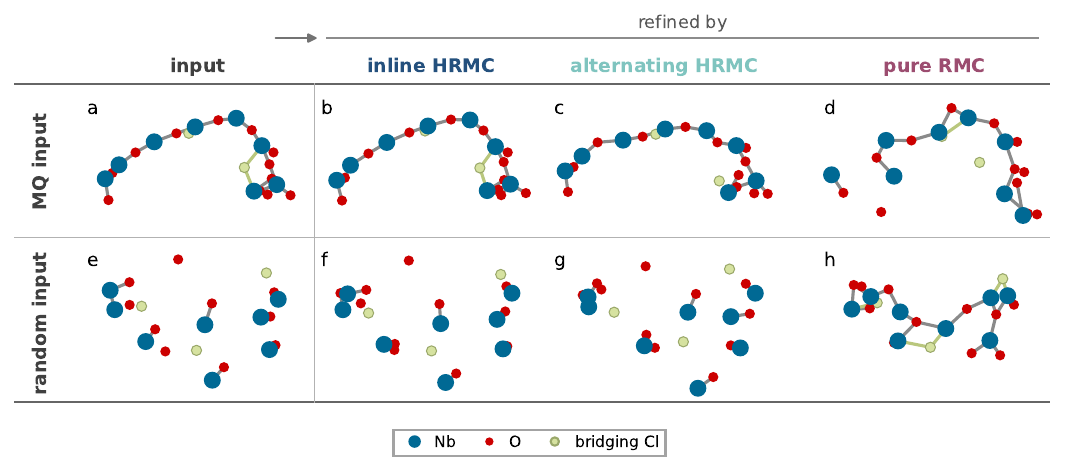}
\caption{One \ce{Nb-O} fragment followed through each route from both inputs. Eight \ce{Nb} and their oxygen neighbors are tracked by atom index. The upper row follows the melt-quench input and the lower row the random input, and the columns show the input, the inline scheme, the alternating scheme, and pure RMC with the coordination-number constraint. Coordinates are unwrapped and projected onto the best-fit plane through the niobium of the row's input, with one projection and one scale per row. Lines mark \ce{Nb-O} contacts below $2.92$~\AA{} and \ce{Nb-Cl} contacts below $3.20$~\AA{} in each panel's own coordinates. 
The fragment shown is the one nearest the median topology statistics of the cell.}
\label{fig:topology}
\end{figure*}

Two choices decide what a hybrid scheme returns. The first is the input, which largely decides the network the refinement ends with.
Figure~\ref{fig:topology} follows one \ce{Nb-O} fragment through each route. Within the tested budgets, both hybrid schemes preserve the input network: the melt-quench chain keeps its links, and the random packing keeps its largely isolated units, whereas pure RMC yields a similar network from either input (Supplementary Table~S1).
The two hybrid networks are hard to distinguish by fit and energy alone: the structures from the two inputs differ by less than $0.02$ in $R^{2}$ and by up to a tenth of the pure RMC penalty in energy. 
Both quantities sum over the whole cell and change little when the polyhedra connect differently. The input is therefore part of the method.

We suggest a melt-quench glass from an available potential, such as a foundation potential, as the input, provided the potential describes the short-range bonding well. The potential then predicts the network rather than assuming it, and the structures refined from the glass lie $0.06$--$0.14$~eV/atom below those refined from the random packing. We chose the r$^2$SCAN head for this reason (Sec.~\ref{sec:methods:mlp}), although we did not compare it with the PBE head.

An MQ input requires a crystal structure of the target composition, which sets the cell and the volume at which the melt is held and quenched (Table~\ref{tab:mq}). For a composition with no reported polymorph, crystal structure prediction by conventional workflows~\cite{pickard_2011, oganov_2006, wang_2010} or by generative models~\cite{zeniGenerativeModelInorganic2025, Luo_2025, zhong_crystal_2025, park_exploration_2025} can supply it. For example, \citet{zhong_machine-learning-guided_2026} generated low-energy polymorphs with a diffusion model to start melt-quench dynamics for amorphous solid-electrolyte interphases and used the same polymorphs as the crystalline reference to assess the stability of the amorphous phase.

The second choice is the hybrid scheme, and in practice it is set by computational cost (Supplementary Sec.~S7). Inline refinement scores both terms on every move and ends closer to the MQ glass in energy than the alternating scheme. Its acceptance rule satisfies detailed balance, so its structures sample a defined distribution. It costs an MLIP call for every move that passes the hard-sphere filter, and its energy still decreases slightly at the end of the budget, so the inline energies reported here are upper bounds (Supplementary Sec.~S5). Alternating refinement carries no energy in its acceptance rule and applies the MLIP only in short relaxations after every $500$ accepted moves. Because the energy acts only through these relaxations, the chain settles at a higher energy, $0.08$ against $0.05$~eV/atom above the glass. It becomes stationary long before its budget ends and costs about a quarter of an inline chain per structure. Pure RMC uses no MLIP and runs on a single processor core, yet it bends the \ce{Nb-O-Nb} bridges (median angle $105^{\circ}$ against $159^{\circ}$ in the glass, Supplementary Table~S1) and creates edge- and face-sharing links at $1.3$--$1.5$~eV/atom above the glass. The inline scheme is the choice when one MLIP call per move is affordable and the tightest energy control is wanted. The alternating scheme is the choice when the budget is limited.

This benchmark has several caveats. The sample is not wholly amorphous. Internal-standard refinement shows that about $55$\% of this composition is amorphous, a fraction that is constant across a stoichiometry series~\cite{newnhamCorrelationCoherenceLength2025}. The measured coherence lengths are $11$--$14$~nm~\cite{newnhamCorrelationCoherenceLength2025}, so a $39$~\AA{} cell cannot hold a crystallite.
Each model configuration is therefore a single amorphous phase compared against the whole-sample scattering, and part of the medium-range oscillation that refinement recovers may come from the crystalline fraction.
The comparison between routes is more informative than the absolute fit of any one of them.
Second, the neutron contrast is uneven. Chlorine supplies $79$\% of the mean coherent scattering length $\langle b\rangle$, whereas Li pairs carry only $6.3$\% of the neutron weight (Eq.~\eqref{eq:wab}).
Randomizing all $292$ lithium positions in a refined structure lowers $R^{2}$ by $0.02$ for the inline result and by $0.004$ for pure RMC (Supplementary Sec.~S6). The Li positions in every model are therefore set mainly by the input, the hard-sphere filter, and (in the hybrid schemes) the MLIP rather than by the data. Any use of these models for lithium pathways inherits this limit.
Finally, pure RMC is run here in its basic form, with a hard-sphere filter and at most one coordination constraint. A production refinement would add further constraints and fit several targets at once, including the density where it is known~\cite{keen1990glasses,wicks1995simultaneous}. We did not test how far such constraints would reduce the ligand loss and energy penalty of pure RMC.

In summary, we have benchmarked MQ, pure RMC, and two hybrid schemes on one amorphous oxychloride electrolyte, against one neutron pair distribution function and one foundation potential, scoring each structure on fit, energy relative to the MQ glass, and local chemistry. MQ dynamics gives the local chemistry at the lowest energy but not the medium-range packing. Pure RMC reproduces the measurement at a cost of more than $1.3$~eV/atom, leaving Nb with about half of its chlorine neighbors and adding edge sharing that a coordination-number constraint does not remove. Starting from the MQ glass, both hybrid schemes reach a close fit near the glass energy while keeping the coordination sphere intact. 
A close fit at low energy does not by itself fix how the polyhedra connect, but the input largely determines the network.
We therefore suggest a practical route for amorphous solid electrolytes: quench a glass with a foundation potential that describes short-range bonding well, then refine it against measured total scattering using a hybrid scheme.
A structure with both local chemistry and medium-range order is the starting point for understanding ionic transport, because the barrier a mobile ion encounters is set by the size and connectivity of the sites it passes through~\cite{adams2005bvs} rather than by the composition alone~\cite{binciMechanismsAlkaliIonic2026}. Such structures let future studies analyze site geometry, polyhedron connectivity, and the free volume between them.

\section*{Acknowledgements} 
This work was primarily funded by the NUS-AISI Joint Research Initiative Fund and supported by computational resources from the NUS High Performance Computing facility (NUSREC-HPC-00001).
The experimental characterization was supported by Shenzhen Guangming Science City Development and Construction Co., Ltd. and its Material Genome Big-Science Facilities Platform; Shenzhen Science and Technology Program (KQTD20200820113047086), Guangdong Provincial Key Laboratory of Energy Materials for Electric Power (2018B030322001), Shenzhen Key Laboratory of Solid State Batteries (SYSPG20241211173726011); Multi-Physics Instrument at the China Spallation Neutron Source (CSNS).
The authors thank Daye Zheng for valuable discussions at AISI.

\section*{Data Availability}
The supporting data and code are available at \url{https://github.com/AM3GroupHub/amsrmc}.

\section*{Author contributions}
P.Z. conceived the idea and designed the research. X.W. performed the research and method development. X.Z., D.B., and B.H. collected the experimental data. 
X.W. and P.Z. wrote the manuscript. P.Z. and J.Z. offered insights throughout the project.

\section*{Competing interests}
The authors declare no competing interests.

\bibliography{references}

\end{document}